\documentclass[twocolumn,showpacs,showkeys,preprintnumbers,amsmath,amssymb]{revtex4}
\usepackage{graphicx}
\usepackage{dcolumn}
\usepackage{bm}

\begin{document}
\preprint{International Journal of Theoretical Physics (IJTP)}
\title{\small Horizon-less Spherically Symmetric Vacuum-Solutions in
a Higgs Scalar-Tensor Theory of Gravity}
\vspace{0.4cm}
\author{Nils M. Bezares-Roder$^1${\footnote{e-mail: Nils.Bezares@uni-ulm.de}}, Hemwati Nandan$^2${\footnote{e-mail:
hntheory@yahoo.co.in}}
 and  Heinz Dehnen$^3${\footnote{e-mail: Heinz.Dehnen
 @uni-konstanz.de}}} \vspace{0.4cm}
\affiliation{$1.$ Institut f$\ddot{u}$r Theoretische Physik,
Universit$\ddot{a}$t Ulm,
Albert-Einstein-Allee 11,$\,D$-$89069$ Ulm, Germany. \\
$2.$ Centre for Theoretical Studies, Indian Institute of Technology,
$721 302$ Kharagpur, India. \\
$3.$ Fachbereich Physik, Universit$\ddot{a}$t Konstanz, Postfach $M
677,\,78457 $ Konstanz, Germany.} \vspace{0.1cm}

{\small{International Journal of Theoretical Physics (IJTP)}\\
DOI 10.1007/s10773-007-9359-5\\
The original publication is available at www.springer.com\\
Int.J.Theor.Phys. \underline{46}(10) 2429-2436 (2007)}

%\date{\today}
\begin{abstract}
\noindent The exact static and spherically symmetric solutions of
the vacuum field equations for a Higgs Scalar-Tensor theory (HSTT)
are derived in Schwarzschild coordinates. It is shown that in general there exists no Schwarzschild horizon and that the fields are only singular (as naked singularity) at the center (i.e. for the case of a pont-particle). However, the Schwarzschild solution as in usual general relativity (GR) is obtained for the vanishing limit of Higgs field excitations.
\end{abstract}
 \maketitle

%\section{The model}
\noindent The standard model (SM) of elementary particle physics has
been remarkably successful in providing the astonishing synthesis of
the electromagnetic, weak and strong interactions of fundamental
particles in nature \cite{Lee,Peskin}, and according to it, any
massive elementary particle is, in fact, surrounded by a scalar
meson cloud represented by the excited scalar Higgs field that acts
as a source of the mass of the particle. The Higgs mechanism
\cite{Higgs,Engelert}, therefore, provides a way of the acquisition
of mass by the gauge bosons and fermions in nature. In view of these
developments, the gravitational field equations with an unknown
additional, ad hoc minimally coupled scalar field added as source in
the Hilbert-Einstein field equations was examined in detail in the
past \cite{Har93}, assuming for simplicity the scalar field as
massless; it was shown that any such scalar field influences and
modifies the metric independently from its strength in such a way
that there exists always a simultaneous solution of the massless
scalar field equation and Einstein's field equations for the static
case with a scalar point-charge as a source. In any case no
Schwarzschild horizon appears and only at the point-particle the
metric and the scalar field result singular as naked singularity;
this is a similar situation to that of the
Reissner-Nordstr$\ddot{o}$m solution \cite{Rei16}, only in a more
general way, since singularities are avoided there only in the case
of sufficiently strong electric charges compared with the mass.\\
\noindent General relativistic models with a scalar field, which is
coupled to the tensor field of GR, are conform equivalent with
more-dimensional general relativistic models \cite{Cot97}. Thus,
following the isomorphy theorem \cite{Jordan}, projective spaces as
Kaluza-Klein's can be reduced to usual Riemannian 4-spaces, whereas
a functional 5th component of the projective metric plays the role
of a variable gravitational ``constant'' (as first worked out in
\cite{Bra61}), by which it is possible to vary the strength of
gravitation \cite{Fau01} in scalar-tensor theories. Following this
and willing to get a Lagrangian-based model from which the
characteristics of the scalar field can be known, in the present
paper, an analogous approach as the one in \cite{Har93} is
fulfilled, but for the case of a non-minimally coupled Higgs field
within the Higgs Scalar-Tensor theory (HSTT); it is shown that this
can lead to regular fields except for the point-particle as naked
singularity.\\
\noindent The interaction of the Higgs field with the particles that
acquire mass is of short-ranged gravitational type
\cite{Deh90a,Deh90b,Deh91} and is also compatible with Dirac's large
number hypothesis \cite{Dirac} and with Einstein's Mach-Principle
\cite{Ein13}, with mass appearing through gravitational interaction.
Utilizing the Jordan-Brans-Dicke theory \cite{Jordan,Bra61} and
Zee's ideas of induced gravity \cite{Zee79}, such Higgs gravitation
was first acquainted by Dehnen and Frommert \cite{Deh92,Deh93},
coupling the Higgs field $\phi$ non-minimally with the curvature
scalar $R=g^{\mu \nu} R_{\mu \nu}$ of General Relativity (GR) and
combining gravity with symmetry breakdown. Such theories have been
subsequently used to explain various diverse physical phenomena viz.
dark matter and flat rotation curves of spiral galaxies
\cite{Ges91,Rod04,Bez06a}, quintessence and cosmological
inflation \cite{Cer1,Cer2,Berkin}.\\
\noindent Let us consider the uniquely formed Lagrangian in the
natural system of units \cite{Deh92} as follows,
\begin{equation}
{\cal L}= \left[\frac{\check{\alpha}}{16\pi}\, \phi^\dagger \phi R
\,+ \, \frac{1}{2} \phi^\dagger_{\, ;\, \mu} \phi^{\,; \, \mu}-
V(\phi)\right] \sqrt{-g}+ {\cal L}_M \sqrt{-g},\label{La}
\end{equation}
where $\check{\alpha}$ is a dimensionless parameter and ${\cal L}_M
\sqrt{-g}$ is the Lagrangian for the fermionic and massless bosonic
fields. The Higgs potential in Eq.(\ref{La}), which is normalized in
such a way that $V(\phi_0)=0$ is valid for the ground state value
$\phi=\phi_0$ of the scalar field, yields
\begin{equation}
V(\phi)= \frac{\mu^2}{2}\phi^\dagger \phi+ \frac{\lambda}{24}
(\phi^\dagger \phi)^2+ \frac{3}{2} \frac{\mu^4}{\lambda} = \frac
{\lambda}{24} \left( \phi \, \phi^\dagger + 6 \,
\frac{\mu^2}{\lambda}\right)^2\,, \label{H}
\end{equation}
with $\mu^2 < 0$ and $\lambda > 0$ as real-valued constants.\\
\noindent The Higgs field in the spontaneously broken phase of
symmetry leads to the ground state value of the Higgs field as
\begin{equation}
\phi_0\,\phi_0^{\dagger} = v^2 \,= \, - \,
\frac{6\mu^2}{\lambda}\,, \label{ground}
\end{equation}
which can further be resolved as $\phi_0=vN$ with $N=const.$
satisfying $N^\dagger N=1$. With the introduction of that unitary
gauge \cite{Deh91,Bez06a}, the general Higgs field $\phi$ may then
be written in terms of the real-valued excited Higgs scalar field
$\xi$ in the following form:
\begin{equation}
\phi=v\sqrt{1+ \xi}\,N \,.\label{phixi}
\end{equation}
The length scale of this Higgs field is given through
\begin{equation}
l= \left[ { \frac{1 + \frac {4 \pi }{3 \breve{\alpha} } } {16 \pi G
( \mu^4 / \lambda ) } } \right]^{1/2}\, = \, M^{-1} \, \label{M}
\end{equation}
with the Higgs field mass $M$, while the gravitational coupling
parameter $G$ is then defined through the ground state value of the
Higgs field given by Eq.(\ref{ground}):
\begin{equation}
G =\frac{1}{\breve{\alpha} v^2} = \,- \,\frac{1}{\breve{\alpha}}\,
\frac{\lambda}{6\mu^2}\, . \label{G}
\end{equation}
The dimensionless parameter $\breve{\alpha}$ in Eq.(\ref{La}) may be
defined in terms of the ratio
\begin{equation}
\breve{\alpha} \simeq (M_{P} / M_{W} )^2\gg 1\,,
\end{equation}
where $M_P$ and $M_W$ are the Planck mass and the mass of the gauge
boson, respectively. The mass of the gauge boson is given through
\begin{equation}
M_{W} = {\sqrt \pi} g v,
\end{equation}
where $g$ is the coupling constant of the corresponding gauge group.
The effective gravitational coupling, however, is given through
\begin{equation}
G_{eff}=G(\xi) = (1+ \xi)^{-1}G \, ,
\end{equation}
which reduces to Eq.(\ref{G}) in the absence of the excited Higgs
scalar field $\xi$ (i.e. for $\xi=0$) and gets singular for a
vanishing Higgs scalar field with $\xi=-1$ \cite{Bez06a}.\\
\noindent The coupling (given through ${\cal L}_M$) of the Higgs
particles to their source is only weak (i.e. $\sim G$)
\cite{Deh92,Bez06a,Bez06H}, and the Higgs field equation takes the
form
\begin{equation}
\xi^{,\,\mu}\,_{\, ; \, \mu}+ \frac {\xi}{l^2} \, = \, \frac{8\pi
\,G}{3} \, T,\label{Higgs1}
\end{equation}
for the first of these cases, wherein $T$ is the trace of the 
symmetric energy-momentum tensor
$T_{\mu \nu}$ belonging to ${\cal{L}}_M \sqrt {-g}$ in the
Lagrangian given by Eq.(\ref{La}), which satisfies the conservation
law $T_\mu\,^\nu\,_{;\nu} =0$ in the here analized case where $\phi$ does not
couple to the fermionic state $\psi$ in ${\cal{L}}_M \sqrt {-g}$.
However, a coupling to SM, which means the production of the
fermionic mass through this Higgs field, breaks the conservation law
through a new ``Higgs force'' and implies simultaneously that the
right hand-side of Eq.(\ref{Higgs1}) vanishes identically
\cite{Deh93,Bez06H}. Nevertheless, this second case will be
discussed later and not further here. Further, the Hilbert-Einstein
equations, which reduce to the  usual ones of GR for vanishing
excitations $\xi$, are now given through {\begin{widetext}
\begin{equation}
R_{\mu \nu}- \frac{1}{2l^2}\left(1+ \xi\right)^{-1} \left[\xi\left(1+
\frac{3}{2}\xi\right)g_{\mu \nu}+ \xi_{,\mu;\nu}\right]\,
=\,-8\pi G_{eff}\left(T_{\mu \nu}- \frac{1}{3}Tg_{\mu \nu}\right),\label{Field}
\end{equation}
\end{widetext}}
\noindent for which the value of the Ricci curvature $R$ has already been
introduced.\\
%\section{Spherical symmetry and limiting case}
\noindent It is important to notice that in view of the structure of
$l$ in the HSTT, only large values of the length scale $l$ are
expected. Indeed, only such values within the HSTT lead to the
correct explanation of flattened rotation curves of spiral
galaxies without assuming dark matter \cite{Ges91,Bez06a}, as a
consequence of long-range changes of the dynamics because of Higgs
gravitation, i.e. the gravitational interaction of massive scalar
Higgs-like fields, where non-minimally coupled Higgs particles
interact almost only through the gravitational channel and thus with
predictions in accordance with the experiments. However, given the
relatively small masses $M$, it is important to notice that the
limiting case of a vanishing Higgs field mass (\ref{M}) of the
non-minimally coupled Higgs field as scalar field ($l\rightarrow
\infty$) can be understood as a double limit $\mu^2\rightarrow 0$
and $\lambda\rightarrow 0$, so that $\mu^4/\lambda=0$ and
$v^2=\mu^2/\lambda=finite$ remain valid throughout. Thus, the ground
state value keeps the degeneracy (remains the one of a Higgs mode
and does not go through to one of a Wigner one) and the symmetry
stays broken at low energies. The scalar field still changes the
usual dynamics after symmetry breakdown and the excitations are in
general non-vanishing. Thus, the field equations (\ref{Field}) do
not reduce to the usual ones of GR and new changes in the dynamics
can be acquainted to the scalar field and its gravitational
Yukawa-type interaction.\\
\noindent An analysis of the limit of vanishing non-minimally
coupled Higgs field masses is important to give general
characteristics of the dynamics within the HSTT, especially if these
are expected as small. Thus, in the following, this Higgs
mass will be neglected and, in order to solve the Eq.(\ref{Higgs1})
with a vanishing Higgs field mass, let us write the line element in
the spherical symmetry ($x^\mu$=\{$x^0=t, x^1=r, x^2=\vartheta,
x^3=\varphi\}$) as
\begin{align}
ds^2=e^{\nu} (dt)^2- e^{\lambda} dr^2- r^2(d\vartheta^2+ sin^2
\vartheta d\varphi^2)\, , \label{element}
\end{align}
where $\nu$ and $\lambda$ are functions of the $r$ coordinate only.
Under the assumption of a point-mass at $r=0$, the Higgs field
equation given by Eq.(\ref{Higgs1}) then takes the form
\begin{equation}
\xi^{''}  \, - \,\frac{1}{2}\,(\lambda^{'} - \nu^{'})\,\xi^{'} +
\frac{2}{r} \, \xi{'} = 0\, ,\label{Higgs2}
\end{equation}
where the prime denotes the differentiation with respect to the
radial coordinate $r$. The first derivative of the excited scalar
field $\xi$ from Eq.(\ref{Higgs2}) in the case of a point-mass (with
internal structure (pressure)) at $r=0$ then reads
\begin{align}
\xi'=\frac{A}{r^2}\, e^{w/2} = \frac{A}{r^2}\,e^{(\lambda -
u/2)}\,,\label{H1}
\end{align}
where
\begin{align}
u:= \lambda+ \nu\quad \text{and}\quad w := \lambda- \nu
\end{align}
are defined. However, the integration constant $A$ is given
according to Eq.(\ref{Higgs1}) with the signature of (\ref{element}) 
in the limit $r\rightarrow \infty$ by
\begin{align}
A= \,-\frac{2}{3} \, G\int T \sqrt{-g} \, d^3 x.\label{a}
\end{align}
With Eq.(\ref{H1}) along with
\begin{align}
q:= \ln(1+ \xi),
\end{align}
the non-trivial field equations associated to the Lagrangian
(\ref{La}) for the metric (\ref{element}) (viz. \cite{Bez06a} for
the complete equations) lead in the case of a point-mass in vacuum
to the following equations:
\begin{equation}
\frac{1}{2}rw'= 1 - e^{(u+w)/2} + r\,q'\, \label{q} ,
\end{equation}
\begin{equation}
u'\, \left(\, 1 + \frac{r}{2}q'\,\right)\, = \, \frac{r}{2}q'\,\left(\,w'  -
\frac{4}{r}\,\right) \, ,\label{u'}
\end{equation}
\begin{equation}
 \frac{1}{2}\,\left(u'-w'\right)=
\frac{B}{r^2}\, e^{w/2- q}=\frac{B}{A}\,q'\, , \label{n'2}
\end{equation}
whereas Eq.(\ref{u'}) is the substraction of field equations and
(\ref{n'2}) is the total integral for $\nu'$ with $B$ as an
integration constant.\\
\noindent Using the value of $u'$ given in Eq.(\ref{u'}), 
Eq.(\ref{n'2}) leads to the following decoupled equation:
\begin{equation}
w'=  -\frac{2(A+B)}{r^2}\, e^{w/2-q}- \frac{AB}{r^3}\,
e^{w-2q}.\label{v'}
\end{equation}
Now, using (\ref{q}) and (\ref{v'}) one immediately deduces
\begin{equation}
e^{u/2+ q} = e^{-w/2+ q}+ \frac{(2A+ B)}{r}+
\frac{AB}{2r^2}\,e^{w/2-q}\,,\label{eu}
\end{equation}
and, therefore, only the differential Eq.(\ref{v'}) remains to be
solved. These considerations further lead to the solution of the
excited Higgs field given by Eq.(\ref{H1}) in the following form for
$B \neq 0$:
\begin{equation}
\xi = -1 + e^q = -1+ e^{\frac{A}{2B}(u-w)}.\label{pnu}
\end{equation}
Eq.(\ref{pnu}) clearly shows that such excitations of the Higgs
scalar field are only possible for a non-vanishing value of the
integration constant $A$ given by Eq.(\ref{a}). The exponential term
with coefficient of amplitude and gravitational potential gives the
deflection from completely vanishing scalar fields.\\
\noindent As boundary condition we postulate the Minkowski metric at
spatial infinity. In order to determine the meaning of the
integration constant $B$ we consider at first the asymptotic case
$r\rightarrow \infty$ of the potentials, i.e. $|w|\ll 1$, $|u|\ll
1$. Then, we get from (\ref{v'}):
\begin{equation}
u=2\frac{A}{r} + \frac{AB}{2r^2}\,,\label{u2}
\end{equation}
\begin{equation}
w= \frac{2(A+B)}{r}+ \frac{AB}{2r^2}\,.\label{v2}
\end{equation}
This results in
\begin{equation}
\nu = \frac{(u-w)}{2} \,= -\frac{B}{r}\quad \, \label{nu1}
\end{equation}
and
\begin{equation}
\lambda=  \frac {(u+w)}{2}\, = \frac{AB}{2r^2}\,  +
\frac{(2A+B)}{r}\,.\, \label{lanu}
\end{equation}
\noindent Consequently,
\begin{equation}
B= \frac {2 \tilde M_S} {\breve{\alpha} v ^2} = 2 \tilde M_S G \,
\label{b}
\end{equation}
is valid in view of the equation of motion of the line element
(\ref{element}), where $\tilde M_S$ is the asymptotic ($r\rightarrow
\infty$) visible mass of the particle (and represents the
Schwarzschild mass).
Further, the differential Eq.(\ref{v'}) is an Abelian one and can be
solved exactly. With the substitution
\begin{equation}
e^{w/2- q}=: r\,\tilde g(r) =: r\,\tilde g,\label{rg}
\end{equation}
Eq.(\ref{v'}) acquires a much simpler form as given below,
\begin{equation}
r\tilde g' = \alpha \tilde g^3- K \, \tilde g^2- \tilde
g\,,\label{g'K}
\end{equation}
whereas
\begin{alignat}{1}
&K:= 2 A + B = 2 (A+G \tilde M_S)\quad  \text{and}\\
&\alpha:=-\frac{AB}{2}=-A\tilde{M}_S G.
\end{alignat}
\noindent Eq.(\ref{g'K}) can be integrated by using the method
of separation of variables, which for $\alpha\neq 0$ reduces to the
form given as
\begin{equation}
\left|\frac{\tilde g^2}{1+ K \tilde g- \alpha \tilde g^2}\right|
\left|\frac{\sqrt{K^2+ 4\alpha}+ K- 2\alpha \tilde g}{\sqrt{K^2+
4\alpha}- K+ 2\alpha \tilde g}\right|^{\frac{K}{\sqrt{K^2+
4\alpha}}}=\frac{C}{r^2}\,. \label{exponen}
\end{equation}
The integration constant $C$ in Eq.(\ref{exponen}) can be calculated
in the Minkowskian limit \cite{Har93} as
\begin{equation}
C=\left(\frac{\sqrt{K^2+ 4\alpha}+ K}{\sqrt{K^2+ 4\alpha}-
K}\right)^\frac{K}{\sqrt{K^2+ 4\alpha}}.\label{K}
\end{equation}
Further, the constant $K$ turns out to be a generalized mass
parameter and $\alpha$ itself can be interpreted as a product-charge
in terms of $A$ and $B$, especially since Eq.(\ref{exponen}) turns
out to be formally the solution for a minimally coupled massless
scalar field added to Einstein's field equations \cite{Har93} (of
course, more than formally only for small values of the $\xi$ field
and thus $q\approx 0$). Thus, the non-minimally coupled massless
Higgs field within the HSTT acts in an ananalogous way to a massless
scalar field within Einstein's theory of gravity in \cite{Har93}.
The integration constants, however, are of different nature (since
$K$ and the charge $\alpha$ are given by both the parameters of the
fields), and a large length scale $l$ is expected, indeed.
Furthermore, symmetry breakdown is still given, since the ground
state stays degenerate and doesn't go through to the one of a Wigner
mode.\\
\noindent In view of the Eqs.(\ref{eu}), (\ref{pnu}) and (\ref{rg}),
the metric components given by Eq.(\ref{element}) and the scalar
field by Eq.(\ref{H1}) for the case $B\neq 0$ may then be
expressed in terms of $\tilde g$ in the following form:
\begin{equation}
\quad e^\nu = \left[\frac{1}{r^2 \tilde g^2}\,(1+ K \tilde g- \alpha
\tilde g^2 )\right]^{B/K} \,,\label{g0}
\end{equation}
\begin{equation}
e^\lambda= 1+ K \tilde g -\alpha \tilde g^2 \,,\label{g1}
\end{equation}
\begin{align}
\quad \xi=\,-1+ \left[\frac{1}{r^2 \tilde{g}^2}(1+ K\tilde{g}- \alpha
\tilde{g}^2)\right]^{\frac{A}{K}}.\label{xie}
\end{align}
The only effective physical parameters remaining in the theory of
the present model are the integration constants $A$ and $B$ defined
by Eqs.(\ref{a}) and (\ref{b}), respectively.\\
%With (\ref{g0}) through (\ref{xie}), using (\ref{g'K}) there is
%\begin{alignat}{1}
%&\nu'=\,-\lambda'\,-p\,(r\,\nu'\,+2)=\, \frac{B}{r}\,\tilde{g}\\
%&\text{and}\,\, \lambda'=\, -\frac{K}{\tilde{g}}\,\tilde{g}+
%2\frac{\alpha}{r}\,\tilde{g}^2,
%\end{alignat}
%so that
%\begin{equation}
%p'\,r\,\nu'=\, -2\frac{\alpha}{r}\,\tilde{g}^2.
%\end{equation}
%These are simple equations of $\tilde{g}(r)$. 
Unfortunately, it is
quite difficult to solve the equation (\ref{exponen}) for $\tilde g$
explicitly. However, a transparent discussion of the properties of
the solution is feasible in connection to \cite{Har93}. For the
limiting case $A=0$, i.e. for the equation of state
$p=\frac{1}{3}\varrho$ (see the Eq.(\ref{a})) and $B\neq 0$ (i.e.
$\alpha=0$ and $K=B$), using Eqs.(\ref{eu}) and (\ref{rg}), Eq.
(\ref{g'K}) can be exactly solved for $\tilde g$ in the following
form (for $B=0$ follows $A=0$ directly and thus the
Minkowski metric with a vanishing value of the excited Higgs field):
\begin{align}
\tilde{g}= \frac{1}{r}\, \left(\,1- \frac{B}{r}
\right)^{-1},\label{g}
\end{align}
and thus for the potentials, using Eqs.(\ref{g0}) and (\ref{g1}):
\begin{align}
e^\nu =\, e^{-\lambda}=\, \left(\,1- \frac{B}{r}\,\right).\label{ln}
\end{align}
This is the Schwarzschild metric, according to Eq.(\ref{element}),
and it appears for the limiting case $A=0$ (i.e. $\xi=0$)\cite{Bez07India}. For
general values of $A$, however, the qualitative course shown in
\cite{Har93} is valid. Moreover, high values of $A$ in (\ref{g0})
lead to a decrease in $\nu$ through the exponent $B/K$. In fact, both the 
metric and scalar field are regular everywhere with exception of r=0 as naked 
singularity. There exists no Schwarzschild horizon, 
which appears only for the case $A=0$ of a vanishing scalar field
excitation. Black holes in the usual sense do not appear in the case
$A\neq 0$. In this case, the scalar field does not lose its special
characteristics and acts anti-gravitationally, canceling a
singularity of curvature and metric except in the center as naked
singularity.
%Higher values of the constant $A$, for instance, lead
%to a smaller gravitational potential, in comparison to the classical
%case.
This is achieved with knowledge of the dynamical equations of
the scalar field and from a Lagrangian (\ref{La}) in the limiting
case of massless fields. Herewith, the scalar field leads to a
screening of the usual gravitational interaction (viz. quintessence
and dark energy). For higher masses, this feature should be also
valid, although in a weakened form.\\
An analysis on flat rotation curves within HSTT \cite{Bez06a}
without dark matter still leads to the possibility of the existence
of highly massive cores for the galaxies. Moreover, these cores, if highly
massive, seem to be necessarily even more massive than within standard 
dynamics. This heavy mass, however, should not be measured as such due to
the characteristics of the scalar field excitation at the galactic
center and to the vanishing of the effective value of the
gravitational coupling. Only the effective value of mass is directly
perceived (which, on the other side, is what may acquaint in the end
for the solution of the dark matter problem) and the effect of the
``real'' central mass is highly screened by the presence of the
scalar field, as would be the case of the gravitative interaction
around this core.\\

%\section{Conclusions}
%The non-minimally coupled scalar field of well-known potential of
%Higgs-type acts gravitational-screening within a model of induced
%gravity so that there exists no Schwarzschild horizon in the limit
%of a vanishing mass of this scalar field coupled to gravity. Only in
%the center (point-particle) the fields are singular as a naked
%singularity. Within this limit, however, the scalar field excitation
%does not vanish and a degenerate ground state of a Higgs mode is
%still given.\\
%For vanishing excitations of the scalar field, though, usual GR is
%recovered and the Schwarzschild solution is re-obtained.

{\small{
\begin{acknowledgments}
\noindent The authors NMBR and HN are thankful to Prof. H. Dehnen and
Prof. P. K. Raina for the hospitality and support during their stay
at the Department of Physics at the University of Konstanz, Germany
and the Centre for Theoretical Studies, Indian Institute of
Technology, Kharagpur, India (under its visitor's programme)
respectively. The authors would also like to thank Prof. F. Steiner
and Dr. R. Aurich of the Institute of Theoretical Physics, 
University of Ulm, for support and useful comments.\\\\

These results were partially presented at the New Delhi IAGRG
meeting, Feb. 2th, 2007.
\end{acknowledgments}}}
\bibliography{apssamp}

\end{document}